\documentclass[prd,amsmath,showpacs,preprintnumbers,
               floatfix,twocolumn,letterpaper,h-physrev4-1,nofootinbib]{revtex4-1}
\usepackage{times}
\usepackage{graphicx}
\usepackage{slashed}
\usepackage{amssymb}
\usepackage{mathrsfs}
\usepackage{epsf}
\newcommand{\nc}{\newcommand}
\nc{\lb}{\langle}
\nc{\rk}{\rangle}
\nc{\Blb}{\Big\langle}
\nc{\Brk}{\Big\rangle}

\nc{\mi}{\!\!\mid\!\!}
\nc{\ra}{\rightarrow}
\nc{\Ra}{\Rightarrow}
\nc {\cd}{\partial}
\nc {\sla}{\slashed}
\nc{\ro}{\mathrm}
\nc{\ca}{\mathcal}
\nc{\sr}{\mathscr}
\nc{\Tr}{\ro{Tr}}
\nc{\Str}{\ro{Str}}
\nc{\realtrace}{\ro{Re\; Tr}}
\nc{\maxrealtrace}{\ro{max\, Re\; Tr}}
\nc{\ud}{\ro{d}}
\nc{\nn}{\nonumber}

\nc{\pb}{\bar{\psi}}
\nc{\p}{\psi}
\nc{\vp}{\vec{\pi}}
\nc{\vap}{\varphi}
\nc{\vt}{\vec{\tau}}
\nc{\si}{\sigma}
\nc{\Si}{\Sigma}
\nc{\g}{\gamma}
\nc{\G}{\Gamma}
\nc{\la}{\lambda}
\nc{\La}{\Lambda}
\nc{\ep}{\epsilon}
\nc{\de}{\delta}
\nc{\De}{\Delta}
\nc{\cL}{\ca{L}}
\nc{\cLe}{\ca{L}_{\ro{eff}}}
\nc {\ti}{\tilde}
\nc{\f}{\frac}
\nc{\da}{\dagger}
\nc{\SU}{\ro{SU}}

\nc{\darrow}{\stackrel{\leftrightarrow}{\cd}}
\nc{\darrows}{\stackrel{\leftrightarrow}{\sla{\cd}}}
\nc{\Darrows}{\stackrel{\leftrightarrow}{\sla{D}}}
\nc {\mpisq}{m_{\pi}^2}
\nc{\Mc}{\stackrel{\circ}{M}}
\nc{\cc}{\stackrel{\circ}{c}}
\nc{\kc}{\stackrel{\circ}{\kappa}}
\nc{\McHB}{\stackrel{\circ}{M}_N^{\ro{HB}}}
\nc{\McB}{\stackrel{\circ}{M}_N^{\ro{B}}}
\nc{\ccHB}{\stackrel{\circ}{c}_1^{\ro{HB}}}
\nc{\ccB}{\stackrel{\circ}{c}_1^{\ro{B}}}
\nc{\kcHB}{\stackrel{\circ}{\kappa}_{\ro{isov}}^{\ro{HB}}}
\nc{\kcB}{\stackrel{\circ}{\kappa}_{\ro{isov}}^{\ro{B}}}
\nc{\gc}{\stackrel{\circ}{g}}
\nc{\fc}{\stackrel{\circ}{f}}
\nc{\hM}{\hat{M}_N}
\nc {\eqb}{\begin{equation}}
\nc {\eqe}{\end{equation}}
\nc {\eqab}{\begin{eqnarray}}
\nc {\eqae}{\end{eqnarray}}

\begin{document}

\title{Limitations of the heavy-baryon expansion as revealed by a pion-mass dispersion relation}

\author{Jonathan M. M. Hall} 

\affiliation{Special Research Centre for the Subatomic Structure of
  Matter (CSSM), Department of Physics, University of Adelaide 5005,
  Australia}

\author{Vladimir Pascalutsa}
\affiliation{Institut f\"{u}r Kernphysik, Johannes Gutenberg Universit\"{a}t, 
Mainz D-55099, Germany}

\preprint{ADP-12-07/T774, MKPH-T-12-01}

\begin{abstract}
The chiral expansion of nucleon properties such as mass,
magnetic moment, and magnetic polarizability are investigated 
 in the framework of chiral perturbation
theory, with and without the heavy-baryon expansion.  
The analysis makes use of a pion-mass
dispersion relation, which is shown to hold in both frameworks. 
The dispersion relation 
allows an ultraviolet cutoff to be implemented 
without compromising the symmetries. 
After renormalization, the leading-order heavy-baryon loops 
demonstrate a stronger 
dependence on the cutoff scale, which results in 
weakened convergence of the expansion. 
This conclusion is tested 
against the recent results of lattice QCD simulations
for nucleon mass and isovector magnetic moment. 
In the case of the polarizability, the situation is even more dramatic as 
the heavy-baryon expansion is unable to reproduce large soft 
contributions to this quantity.  
Clearly, the heavy-baryon expansion is not suitable for every quantity. 
\end{abstract}

\maketitle

\tableofcontents

\section{Introduction}
\label{sect:intro}

The dependence of hadron properties on the quark masses
--- the chiral behavior --- is crucial for interpreting the modern lattice
 QCD calculations, 
which usually require an extrapolation in
the quark mass. It is also important for determining  
 the quark mass values, 
as well as for any quantitative description of
chiral symmetry breaking.
Chiral perturbation theory ($\chi$PT) \cite{Weinberg:1978kz,Gasser:1983yg}, 
a low-energy effective field 
theory (EFT) of QCD, 
should in principle describe the nonanalytic dependencies
on the light-quark masses in a systematic
fashion. 
The analytic (series-like) dependencies
are more arbitrary in $\chi$PT, and are specified in terms of 
low-energy constants (LECs). They are to be fixed by matching to the 
underlying
theory, which is usually achieved by fitting to lattice QCD results.

In the baryon sector of $\chi$PT one often invokes an additional
expansion in the inverse baryon masses called the heavy-baryon expansion,
or HB$\chi$PT \cite{Jenkins:1990jv,Bernard:1995dp}. 
The baryon $\chi$PT without the heavy-baryon
expansion will be referred to as B$\chi$PT. 
While both
expansions should converge to the same result, the question
is whether they converge in a natural way (cf.~\cite{Georgi:1991ch,Georgi:1994qn} for an explanation of `natural').  If the difference between them
at any finite order is as crucial as is claimed, it is clear that they cannot
both converge naturally. 
Therefore, the goal is to establish which of HB- or B$\chi$PT has the more 
natural expansion. 

Up to a given chiral order, 
the B$\chi$PT result 
can be written as the HB$\chi$PT result
and a series of contributions, which are nominally of
 higher order in HB$\chi$PT. 
Whether these 
contributions
are indeed of smaller size 
can only be checked in explicit calculations,
and many cases it has been observed that the
difference between a given-order of HB- and B$\chi$PT 
results is unnaturally large  \cite{Becher:1999he,Fuchs:2003qc,Pascalutsa:2004ga,Holstein:2005db,Pascalutsa:2005ts,Geng:2008mf,Ledwig:2011cx,Alarcon:2011zs}. 
A notable example is
provided by the magnetic polarizability of the proton, for which the leading
chiral-loop contribution predicts $-1.8$ in B$\chi$PT \cite{Bernard:1991rq}  
or $+1.3$  in HB$\chi$PT \cite{Bernard:1995dp}, in units of
$10^{-4}$ fm$^3$. 
These large differences are reconciled in practice by adopting unnaturally
large values for some of the LECs, appearing at higher orders in HB$\chi$PT
(see e.g.~\cite{Hildebrandt:2003fm}). This is not a solution of course, 
but rather a restatement of the problem.   

\begin{figure}[t]
\includegraphics[height=120pt]{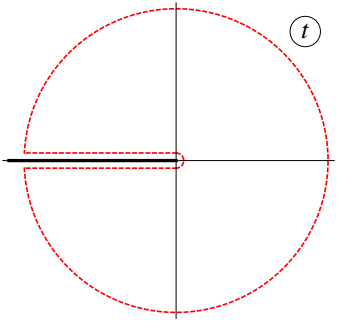}
\caption{(color online). 
The complex $t=m_\pi^2$ plane, with the branch-cut along the
negative real axis, and the contour indicating the analyticity domain.}
\label{fig:cont1}
\end{figure}

In this work, the problem is investigated 
by using the recently established pion-mass dispersion relation 
\cite{Pascalutsa:2009wu,Ledwig:2010nm}: 
\eqb
\label{eqn:disprel}
 f(m_\pi^2) = -\f{1}{\pi}\!\int\limits_{-\infty}^0\!\ro{d}t \,
\f{\ro{Im} \, f(t)}{t-m_\pi^2}\, ,
\eqe
where the static quantity $f$ is a complex function 
of the pion mass squared $m_\pi^2$. Throughout this
work, the focus is on the following static properties of the nucleon: 
the mass $f\equiv M_N$, 
anomalous magnetic moment (AMM) $f\equiv \kappa_N$,
and magnetic polarizability $f\equiv \beta_N$. 
The dispersion relation results from 
the observation that chiral loops in $\chi$PT are analytic functions
in the entire complex plane of $m_\pi^2$ except for the negative real-axis, 
which contains a branch cut associated with pion production, 
see Fig.~\ref{fig:cont1}. 

As will be demonstrated in 
Sec.\ \ref{sect:Bchipt}, there is a correspondence between 
integrating over $t$ in the above dispersion relation and the integration
over the momentum in a chiral loop. 
One can 
cut off the $t$-integration without danger of compromising the symmetries
of the theory.  
Thus an ultraviolet
cutoff $\La$ is introduced in Sec.\ \ref{sect:cutoff}. 
After subtracting positive
powers of $\La$ into the available LECs,  
the aim is to determine the scales
at which differences between the HB- and B$\chi$PT appear. 
It is observed that cutting off the dispersion relation 
is equivalent to the `sharp cutoff' version of 
finite-range regularization (FRR), introduced originally
to improve the HB$\chi$PT expansion by resumming the chiral series through 
the introduction of a regulator into the loop integrals, see
\cite{Donoghue:1998bs,Leinweber:1998ej,Young:2002cj,Young:2002ib,Borasoy:2002jv,Leinweber:2003dg,Bernard:2003rp,Young:2004tb,Leinweber:2004tc,Leinweber:2005xz}.
The equivalence between the cutoff dispersion relation and FRR readily allows 
for an extension of the FRR to the 
realm of B$\chi$PT.

In Sec.\ \ref{sect:calcs},  
the HB- and B$\chi$PT FRR formulae for the cases of
the nucleon mass, AMMs, and magnetic polarizability are obtained at order $p^3$,
 and their residual cutoff-dependence is studied.
In Sec.\ \ref{sect:data}, these formulae are confronted with experimental
and lattice QCD results. The recent lattice QCD simulations   
of PACS-CS \cite{Aoki:2008sm} and 
JLQCD \cite{Ohki:2008ff} Collaborations are used for the nucleon mass, and  
simulations from QCDSF \cite{Collins:2011mk} Collaboration are used for the
 AMMs. 
The conclusions are presented in Sec.\ \ref{sect:conc}. In the Appendix, 
the  dispersion relation in the quark mass is considered briefly, and 
a condition for its  `subtractions' is discussed.
  
\section{Integration over loop momentum vs.\ pion mass}
\label{sect:Bchipt}

\begin{figure}[b]
\centering
\includegraphics[height=65pt]{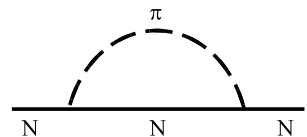}
\vspace{-3mm}
\caption{The pion loop contribution to the self-energy
    of the nucleon, providing the leading nonanalytic contribution to
    the nucleon mass.  All charge conserving transitions are implicit.}
\label{fig:nucSEpiN}
\end{figure}

Consider the example of the nucleon mass, which, 
takes the following simple form 
to order $p^3$ in HB$\chi$PT:
\eqb
\label{eqn:MNexpsn1}
M_N = \, \Mc_N - \, 4 \stackrel{\circ}{c}_1\! m_\pi^2 + \chi_N  m_\pi^3. 
\eqe
$\Mc_N$ and  $\stackrel{\circ}{c}_1$ are the LECs to this order, and 
the coefficient of the nonanalytic term is given by:
\eqb
\chi_N = -\f{3g_A^2}{32\pi f_\pi^2},
\eqe
for the empirical values of $g_A \simeq 1.27$ and $f_\pi \simeq 92.4$ MeV.
The nonanalytic term follows from the self-energy graph in 
Fig.\ \ref{fig:nucSEpiN}, which yields the familiar form in HB$\chi$PT:
\eqb
\label{eqn:HBloop}
\Si_N^{\ro{HB}} (m_\pi^2)= \chi_N\f{2}{\pi}\int\limits_0^\infty\ud k \, \f{k^4}{k^2 + m_\pi^2},
\eqe
where $k$ is the  
magnitude of the 3-momentum running in the loop.
After a change of integration variable to $\,t =-k^2$, 
one obtains:
\eqb
\Si_N^{\ro{HB}} (m_\pi^2) = -\f{1}{\pi}\!\int\limits_{-\infty}^0\!\ud t
\, \f{ \chi_N (-t)^{3/2}}{t - m_\pi^2}.
\eqe
which is simply the
pion-mass dispersion relation of Eq.\ (\ref{eqn:disprel}),
with: $\,\ro{Im}\,\Si_N^{\ro{HB}}(t)  = \chi_N \, (-t)^{3/2}$.
Thus it is not only evident that the heavy-baryon loop obeys the pion-mass
dispersion relation, but also that there is a 
correspondence between the integration
over the pion mass and the loop momentum. It can easily be 
 seen that cutting off the loop momentum at some scale $\La$ is
equivalent to placing the lower-limit of integration over $t$ at $-\La^2$.

Having observed this equivalence, 
the dispersion relation will be used in preference 
to the loop integrals themselves. The reason for this is twofold. 
Firstly, 
in order to express the loop integral 
with only a single $k$-integration,  the $k_0$ 
 and angular integrations must be evaluated, 
which is usually more difficult than finding the imaginary part of the loop 
(especially for multi-loop graphs and B$\chi$PT expressions).
Secondly, 
 it is explicitly 
clear that no symmetries of the theory are compromised in the cutoff 
of $t$. Of course, 
it has been shown that FRR schemes with a sharp cutoff in the momentum 
are consistent with chiral symmetry \cite{Bernard:2003rp}. 
However, in $t$-integration, consistency with 
the symmetries is more explicit, and violations of analyticity 
become apparent immediately.

The cutoff-dependence of a given quantity is meaningful only after 
 the loop contributions have been renormalized. 
When using the
 dispersion relation, the usual procedure of dimensional regularization 
and cancellation of infinities by counter-terms 
 is replaced by `subtractions'. For instance, the nucleon mass 
 requires at least two subtractions at $m_\pi^2=0$ 
(cf.\ Appendix \ref{sect:subt}), 
 so that 
 the third-order contribution can be written as: 
 \eqb
\label{eqn:MNfromdisprel}
M_N^{(3)}
= -  \f{1}{\pi}\!\int\limits_{-\infty}^0\! \ro{d}t\,
\f{\ro{Im} \,M_N^{(3)}(t)}{t-m_\pi^2}\, \frac{m_\pi^4}{t^2} \stackrel{\ro{HB}}{=}
\chi_N \, m_\pi^3,
\eqe
with the full result, to order $p^3$, given by Eq. (\ref{eqn:MNexpsn1}). 
The LECs play the role of the subtraction constants.

Introduction of an ultraviolet cutoff in the finite integral after
 the 
subtractions allows one to separate the low- 
and high-momentum contributions. It can also provide information about 
the scale at which the
HB- and B$\chi$PT results start to deviate.

\section{Introducing a cutoff: finite-range regularization}
 \label{sect:cutoff}

The chiral expansion of an observable quantity $f$ 
is an expansion in the quark mass 
$m_q$ around the chiral limit ($m_q~\to~0$), which in $\chi$PT becomes
an expansion in $p=m_\pi/\La_\chi$, the mass of the 
pseudo-Goldstone boson of spontaneous chiral 
symmetry breaking over the scale of chiral symmetry breaking 
$\La_\chi \simeq 4\pi f_\pi \approx 1$ GeV \cite{Manohar:1983md}.
Because of the branch cut in the complex-$m_\pi^2$ plane 
along the negative real-axis, the chiral expansion is not a series
expansion (otherwise, it would have a zero radius of convergence), but rather 
an expansion in non-integer powers of $m_\pi^2 \propto m_q$.

By writing the dispersion integral as:
\eqb
\label{eqn:disprel5}
f(m_\pi^2) = -\f{1}{\pi}\left( \, \int\limits_{-\La_\chi^2}^0 
+ \int\limits_{-\infty}^{-\La_\chi^2}\, \right) \ro{d}t \,
\f{\ro{Im}\, f(t)}{t-m_\pi^2} ,
\eqe
it is evident that the second integral can  be expanded in
integer powers of $
m_\pi^2/\La_\chi^2
$. Hence this term is of analytic form and can only affect the values of the
LECs. Indeed, the physics above the scale $\La_\chi$ is not
described by $\chi$PT and therefore its effect should be absorbable 
by the LECs.

The second integral generates an infinite number
of analytic terms, while the number of LECs to a given order of the calculation 
is finite. The higher-order analytic terms are present and not compensated
by the LECs at this order, but their effect should not exceed the uncertainty 
in the calculation due to the neglect of all the other higher-order terms.
That is, the second integral can be dropped, while the resulting
cutoff-dependence represents the uncertainty due to higher-order effects.

One purpose of imposing a cutoff of order of $1$ GeV is to 
investigate the convergence of the expansion  
without actually computing any of the higher-order contributions.
This is one the main goals of FRR-- 
indeed, motivated by the absence of rapid curvature in all hadronic 
observables at larger quark masses, FRR aims to resum the chiral expansion 
in the expectation of improving the convergence of the residual series 
\cite{Young:2002ib,Leinweber:2003dg,Leinweber:2005xz,Hall:2010ai,Hall:2011en}.
From the formulae shown in Eqs.~(\ref{eqn:HBloop}) -- (\ref{eqn:MNfromdisprel}) 
in the previous section, it is clear that the 
`sharp cutoff' FRR is equivalent to the cutoff pion-mass dispersion relation:
\eqb
\label{eqn:disprel3}
 f(m_\pi^2; \La^2) = -\f{1}{\pi}\!\int\limits_{-\La^2}^0\!\!\ro{d}t \,
\f{\ro{Im} \, f(t)}{t-m_\pi^2} \left(\f{m_\pi^2}{t}\right)^n,
\eqe
 where $n$ indicates the number of subtractions around the chiral limit.
 In this work,  
the main aim is to see at which values of 
the cutoff any deviation occurs between the  
HB- and B$\chi$PT results.   
If the deviation begins at $\La \ll 1$ GeV, then the differences between 
the two expansions cannot be reconciled in a natural way. 
In the next section, this situation is examined using 
several specific examples, and for each of them a different picture is 
obtained (cf.~Fig.~\ref{fig:LaDep}). 

\section{Nucleon properties at $\ca{O}(p^3)$}
 \label{sect:calcs}
 
At chiral order $p^3$, the imaginary parts of the nucleon mass,
the proton
and neutron AMMs, and the magnetic polarizability of the 
proton were computed in 
Ref.\ \cite{Ledwig:2010nm}\footnote{The original expressions
of \cite{Ledwig:2010nm} contain misprints:
Eqs.\ (10)--(12) miss an overall factor of $4$, while Eq.~(14) misses a factor 
of $\tau$.}:
\begin{subequations}
\begin{align}
\label{ImMN}
\mathrm{Im}\, M_N^{(3)} (t) & =  
 \frac{3 g_A^2 \hat M_N^3}{(4\pi f_\pi)^2} \frac{\pi \tau}{2} 
\Big( \f{1}{2} \tau +\la \Big)\, \theta(-t)\,,\\
\label{eqn:Imkp}
\ro{Im}\, \kappa_p^{(3)} (t)&=
\frac{g_A^2 \hat M_N^2 }{(4\pi f_\pi)^2}  \frac{2 \pi  }{\la}  \Big(\f{1}{2}\tau
 + \la \Big)^2
\Big[ 1-\f{3}{2} \Big( \f{1}{2} \tau +\la \Big) \Big]\nn\\
&\qquad\times \, \theta(-t)\, , \\
\label{eqn:Imkn}
\ro{Im}\, \kappa_n^{(3)} (t)&=
-\frac{g_A^2 \hat M_N^2 }{(4 \pi f_\pi)^2}  \frac{2 \pi  }{ \la}  \Big(\f{1}{2}
 \tau + \la \Big)^2\, \theta(-t)\,, \\
\mathrm{Im}\, \beta^{(3)}_p (t) & = - 
\frac{(e^2/4\pi) \, g_A^2}{(4 \pi f_\pi)^2 \hat M_N }
\frac{ \pi \tau }{24\la^3} \Big[ 2-72 \la \nn\\
& + (418\la-246) \,\tau - (316\la-471) \,\tau^2 \nn\\
& +(54\la-212)\,\tau^3 +27\tau^4\Big] \,\theta(-t),
 \end{align}
  \end{subequations}
 where $\hat M_N\simeq 939$ 
 MeV is the physical nucleon mass,  $e^2/4\pi \simeq 1/137$ is 
 the fine-structure constant, and
 the following dimensionless variables are introduced:
 \eqb
 \tau = \f{t}{\hat M_N^2}, \quad \la = \sqrt{\f{1}{4} \tau^2 -\tau}\,.
 \eqe
  
  The expression for the mass comes from the graph 
  in Fig.~\ref{fig:nucSEpiN}, with leading (pseudo-vector) $\pi NN$
  coupling. The expressions for the AMMs and polarizability
  come from graphs obtained from  Fig.~\ref{fig:nucSEpiN}
  by minimal insertion(s) 
  of $1$- and $2$-photons, respectively.

  The corresponding heavy-baryon expressions at order $p^3$ can be
   obtained by keeping only the leading in $1/\hat M_N$ term 
(i.e, $\la \approx \sqrt{-\tau}$, etc.):
 \begin{subequations}
\begin{align}
\mathrm{Im}\, M_N^{(3)} (t) & \stackrel{\ro{HB}}{=}  
 \frac{3 g_A^2 \hat M_N^3}{(4\pi f_\pi)^2} \frac{\pi \tau}{2} 
\sqrt{-\tau} \, \theta(-t)\,,\\
\label{eqn:HBImkp}
\ro{Im}\, \kappa_p^{(3)} (t)& \stackrel{\ro{HB}}{=}
\frac{ g_A^2 \hat M_N^2 }{(4\pi f_\pi)^2}\, 2 \pi \sqrt{-\tau}
\, \theta(-t) \stackrel{\ro{HB}}{=} - \, \ro{Im}\, \kappa_n^{(3)} (t), \\
\mathrm{Im}\, \beta^{(3)}_p (t) & \stackrel{\ro{HB}}{=} 
\frac{(e^2/4\pi) \, g_A^2}{(4 \pi f_\pi)^2 \hat M_N }
\frac{ \pi }{12\sqrt{-\tau}}  \,\theta(-t).
 \end{align}
 \label{eqn:HBImMN}
\end{subequations}

The full, renormalized result for a given quantity is obtained by substituting
these imaginary parts into the dispersion relation of Eq.~(\ref{eqn:disprel3}). 
The number of subtractions required in each case differ: $n=2$ for $M_N$,
$n=1$ for AMMs, and no subtractions for polarizability.  
The resulting expressions read as follows:

 \begin{widetext}
\begin{subequations}
 \begin{eqnarray}
 \label{eqn:MNfitrel}
M_N (m_\pi^2; \La^2) & = & \, \Mc_N\! - \, 4 \stackrel{\circ}{c}_1\! m_\pi^2 
-\f{\chi_N}{\pi}\!\!\!\int\limits_{-\La^2}^0\!\!\ro{d}t 
\f{\Big[
(-t)^{3/2}\sqrt{1-\f{t}{4\hM^2}} - \f{t^2}{2\hM}\Big]}
{t-m_\pi^2} \left(\f{m_\pi^2}{t}\right)^2 \nn\\
& = & \, \Mc_N \! - \, 4 \stackrel{\circ}{c}_1\! m_\pi^2 
+ \f{\chi_N m_\pi^4}{2\pi\hM}\Bigg\{2\sqrt{\f{4\hM^2}{m_\pi^2}-1}\,
\arctan\Bigg(\f{\La}{m_\pi} \sqrt{\f{4\hM^2-m_\pi^2}{4\hM^2
+ \La^2}}\Bigg)\nn\\
& &+ \,2\,\ro{arcsinh}\f{\La}{2\hM}
+ \log\f{m_\pi^2}{m_\pi^2 + \La^2}\Bigg\}, \\
\label{eqn:frrtildeprot}
\kappa_p(m_\pi^2; \La^2) &=& \stackrel{\circ}{\kappa}_p   + 
\,  \f{g_A^2 \hM^2}{(4\pi f_\pi)^2}
\bigg\{ \f{m_\pi(-8+22\mbox{$\frac{m_\pi^2}{\hM^2}$} - 6
\mbox{$\frac{m_\pi^4}{\hM^4}$})}{\hM\Big( 4-\mbox{$\frac{m_\pi^2}{\hM^2}$} \Big)^{1/2}} \,\arctan\Bigg(\f{\La}{m_\pi} \sqrt{\f{4\hM^2-m_\pi^2}{4\hM^2
+ \La^2}}\Bigg) \nn\\
&-& \f{m_\pi^2}{\hM^2} \Big(5-  \f{3 m_\pi^2}{\hM^2}\Big)
\,\left[ 2\, \ro{arcsinh}\f{\La}{2\hM} +
 \log\f{m_\pi^2}{m_\pi^2 + \La^2}\right] + 
  \f{3m_\pi^2  \La^2}{\hM^4} \left(1-\sqrt{1+\frac{4\hM^2}{\La^2}}\,\right)\bigg\}, \\
%
\label{eqn:frrtildeneut}
\kappa_n(m_\pi^2; \La^2) &= & \stackrel{\circ}{\kappa}_n   + 
 \, \f{4 g_A^2 \hM^2}{(4\pi f_\pi)^2}
\bigg\{ \f{m_\pi(2-\mbox{$\frac{m_\pi^2}{\hM^2}$})}{\hM\Big( 4-\mbox{$\frac{m_\pi^2}{\hM^2}$} \Big)^{1/2}} \,\arctan\Bigg(\f{\La}{m_\pi} \sqrt{\f{4\hM^2-m_\pi^2}{4\hM^2
+ \La^2}}\Bigg) \nn\\
&+& \f{m_\pi^2}{2\hM^2} \left[ 2\,\ro{arcsinh}\f{\La}{2\hM}
+  \log\f{m_\pi^2}{m_\pi^2 + \La^2}\right]\bigg\}, \\
\label{eqn:frrtildebetap}
\beta_p(m_\pi^2; \La^2) &= & \frac{(e^2/4\pi) \, g_A^2}{3 (4 \pi f_\pi)^2 } \Bigg\{ \,\f{2(2-246\mbox{$\frac{m_\pi^2}{\hM^2}$} +471
\mbox{$\frac{m_\pi^4}{\hM^4}$}-212\mbox{$\frac{m_\pi^6}{\hM^6}$}
+27\mbox{$\frac{m_\pi^8}{\hM^8}$})}{m_\pi\Big( 4-\mbox{$\frac{m_\pi^2}{\hM^2}$} \Big)^{3/2}} \,\arctan\Bigg(\f{\La}{m_\pi} \sqrt{\f{4\hM^2-m_\pi^2}{4\hM^2
+ \La^2}}\Bigg) \nn\\
&-& \Big(\f{9}{\hM}-  \f{50 m_\pi^2}{\hM^3} +\f{27 m_\pi^4}{\hM^5}\Big)
\,\left[ 2\, \ro{arcsinh}\f{\La}{2\hM} +
 \log\f{m_\pi^2}{m_\pi^2 + \La^2}\right] \\
 &-& 
  \f{\La^2}{\hM^3} \left[ \f{27(\La^2 - 2 m_\pi^2)}{2\hM^2}
  \left(1-\sqrt{1+\frac{4\hM^2}{\La^2}}\,\right) + 50 - 23 \sqrt{1+\frac{4\hM^2}{\La^2}}  - \f{51 \hM^6}{\La^2 (4\hM^2+\La^2)(4\hM^2-m_\pi^2) }\right]\Bigg\} \,. \nn
\end{eqnarray}
\label{eqn:Bfrr}
 \end{subequations}
\end{widetext}

The heavy-baryon expressions can be
obtained by picking out the leading in $1/\hat M_N$ term, or equivalently, 
by substituting the corresponding imaginary parts from Eq.~(\ref{eqn:HBImMN}),
into the dispersion relation. In the latter case, 
the same integral is encountered in all of the examples:
\begin{align}
\label{eqn:Jeqn}
J(m_\pi; \La)\equiv \int\limits_{-\La^2}^0\!\ro{d}t \, \f{1}{(t-m_\pi^2)\sqrt{-t}}
 =  -\f{2}{m_\pi}\arctan\f{\La}{m_\pi}.
\end{align}
All of the above quantities to $\ca{O}(p^3)$ in HB$\chi$PT are given by this 
integral, up to an overall constant, and a factor of $m_\pi^{2n}$. 
$n$ is the number of subtractions (or pertinent LECs) at this order.  

\begin{figure}[t]
\begin{center}
\includegraphics[height=1.75\hsize]{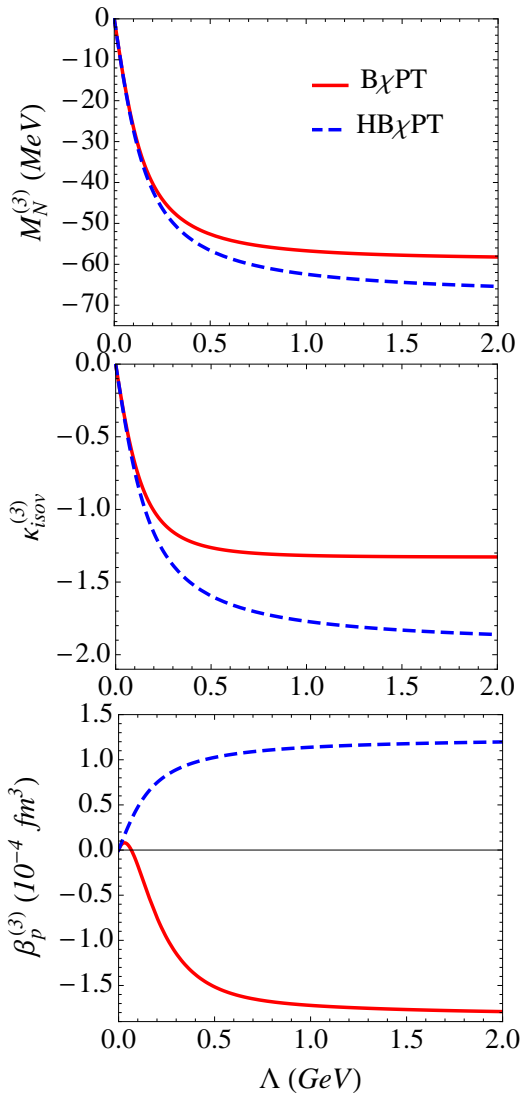}
\caption{(color online).
 The cutoff-dependence of leading-order loop 
contributions to various nucleon quantities (mass, isovector AMM, and
proton's magnetic polarizability) calculated in HB$\chi$PT 
(blue dashed curves) and B$\chi$PT (red solid curves). 
}
\label{fig:LaDep}
\end{center}
\end{figure}

In Fig.~\ref{fig:LaDep}, the resulting cutoff-dependence of 
the above loop contributions is shown at the physical value of the pion mass: 
$m_\pi \simeq 139$ MeV. Each quantity (mass, isovector AMM, and 
polarizability) is presented in a separate
panel, where the results with and without the heavy-baryon expansion
are displayed.
This figure illustrates
the following two features: 
\begin{enumerate}
\item  
As seen from Eqs.~(\ref{eqn:Bfrr}) and (\ref{eqn:Jeqn}), 
the residual cutoff-dependence in HB$\chi$PT 
falls off as $1/\La$ in all of the considered examples, while 
the dependence in the case of B$\chi$PT behaves as 
$ 1/\La^2$ for $M_N$, and as  $ 1/\La^4$ for both AMMs and $\beta_p$
despite the presence of the positive powers of $\La$ in 
Eqs.~(\ref{eqn:Bfrr}b -- d). 
The HB$\chi$PT results have a stronger cutoff-dependence than 
the B$\chi$PT results, indicating a larger impact of the unknown 
high-energy physics to be renormalized by higher-order LECs.
Hence, the HB results are at bigger risk of producing
a large contribution from the high-momentum region. 
Note that, although 
not necessarily immediately apparent from Fig.~\ref{fig:LaDep}, 
the cutoff-dependence of the relativistically improved chiral formulae 
can be obtained from Eqs.~(\ref{eqn:Bfrr}a -- d), and in the heavy-baryon case, 
from the chiral formulae from the FRR literature; e.g. see 
Refs.~\cite{Leinweber:2005xz,Young:2004tb}.  

\vspace{2mm}

\item The  HB- and B$\chi$PT results are guaranteed to be the same at 
small values of $\La$ (and not only at $\La=0$), as can be seen by taking derivatives of 
Eq.~(\ref{eqn:disprel3}) 
with respect to $\La^2$, at $\La=0$. However, at finite values of 
$\La$ the differences
are appreciable. Observing significant differences for $\La$ of order $m_\pi$, 
as in the case of $\beta_p$, 
indicates that the size of the $1/\hM$ terms 
is largely underestimated in HB$\chi$PT.
\end{enumerate}

\medskip

\section{Matching: confronting lattice results}
 \label{sect:data}
 
Eventually, 
the $\chi$PT results must be matched to the underlying theory --
or in practice -- fitted to experimental and lattice QCD simulation results.
In this section, it will be demonstrated that chiral extrapolations based 
on the relativistic 
expressions of Eqs.~(\ref{eqn:Bfrr}a -- d)
 are more stable with respect to cutoff variation. 

The case of magnetic polarizability is interesting. Since there are no
unknown LECs at leading order, it constitutes a genuine prediction. 
Unfortunately,
there are no lattice results for this quantity, while the experimental
data are largely uncertain (see Ref.~\cite{Lensky:2009uv} 
for a recent discussion).  
One thing that the data indisputably show is that $\beta_p$
is small compared to the electric polarizability, and positive, 
which seems to be more consistent with the HB$\chi$PT result.
However, it is well known that $\beta_p$ must have a large positive
contribution from the excitation of the $\Delta(1232)$ resonance 
\cite{Lensky:2009uv}, which
can only be accommodated if the chiral loops are negative and partially 
cancel it out. 

For the nucleon mass, one does not expect much difference between HB- and
B$\chi$PT around the physical pion mass, based on 
Fig.~\ref{fig:LaDep}. For larger
pion masses, however, the difference becomes significant,
and may affect the fit to lattice results as is shown in what follows.

In Figs.~\ref{fig:Aokiextlam} and \ref{fig:extlam}, 
 chiral extrapolations of recent lattice results from   
  PACS-CS \cite{Aoki:2008sm} and JLQCD \cite{Ohki:2008ff} are presented.   
 The different panels correspond to different values of the cutoff $\La$,
while the dashed and solid curves correspond to HB- and B$\chi$PT
fit at order $p^3$. The values of the fit parameters obtained 
using PACS-CS and JLQCD results are shown in 
Tables~\ref{table:MNfitparamsPACS-CS} and \ref{table:MNfitparamsJLQCD}, 
respectively.

 The PACS-CS results were generated using non-perturbatively 
$\ca{O}(a)$-improved
 Wilson quark action at a lattice box size of $\sim 2.9$ fm, but the set 
only contains five points, and there is 
a large statistical error in the smallest 
 $m_\pi^2$ point.
The JLQCD results were generated using  
 overlap fermions in $N_f=2$ QCD. The lattice box length for each
simulation result is $\sim 1.9$ fm, 
with a corresponding lattice spacing is $0.118$ fm. 
The box size is small compared to that of the PACS-CS simulations, 
but the statistical uncertainties in each point are also smaller. 
For simplicity, the fits also neglect possible finite-volume corrections.

 \begin{widetext}

\begin{figure}[t]
\begin{center}
\includegraphics[height=0.33\hsize,angle=90]{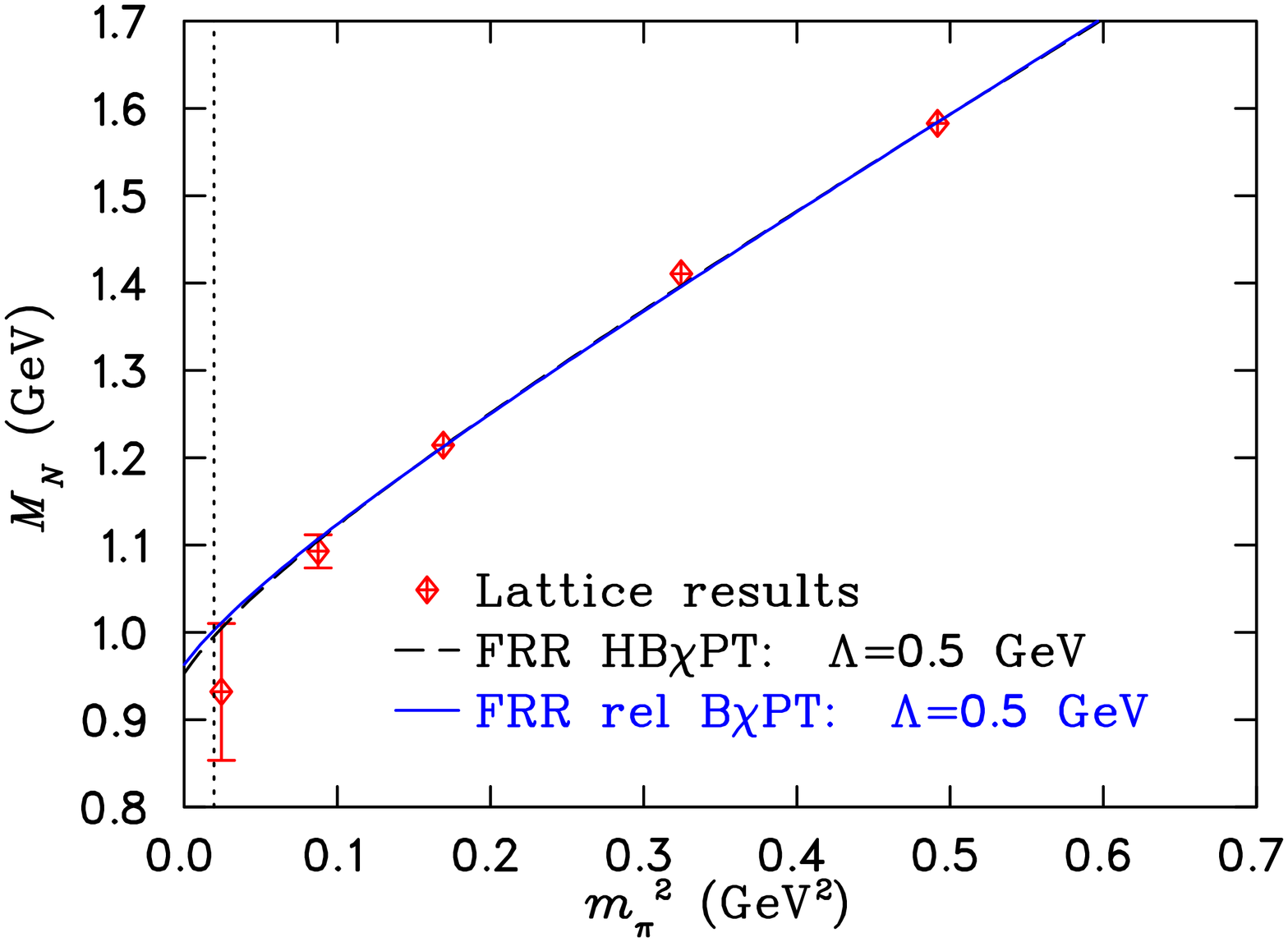}
\includegraphics[height=0.33\hsize,angle=90]{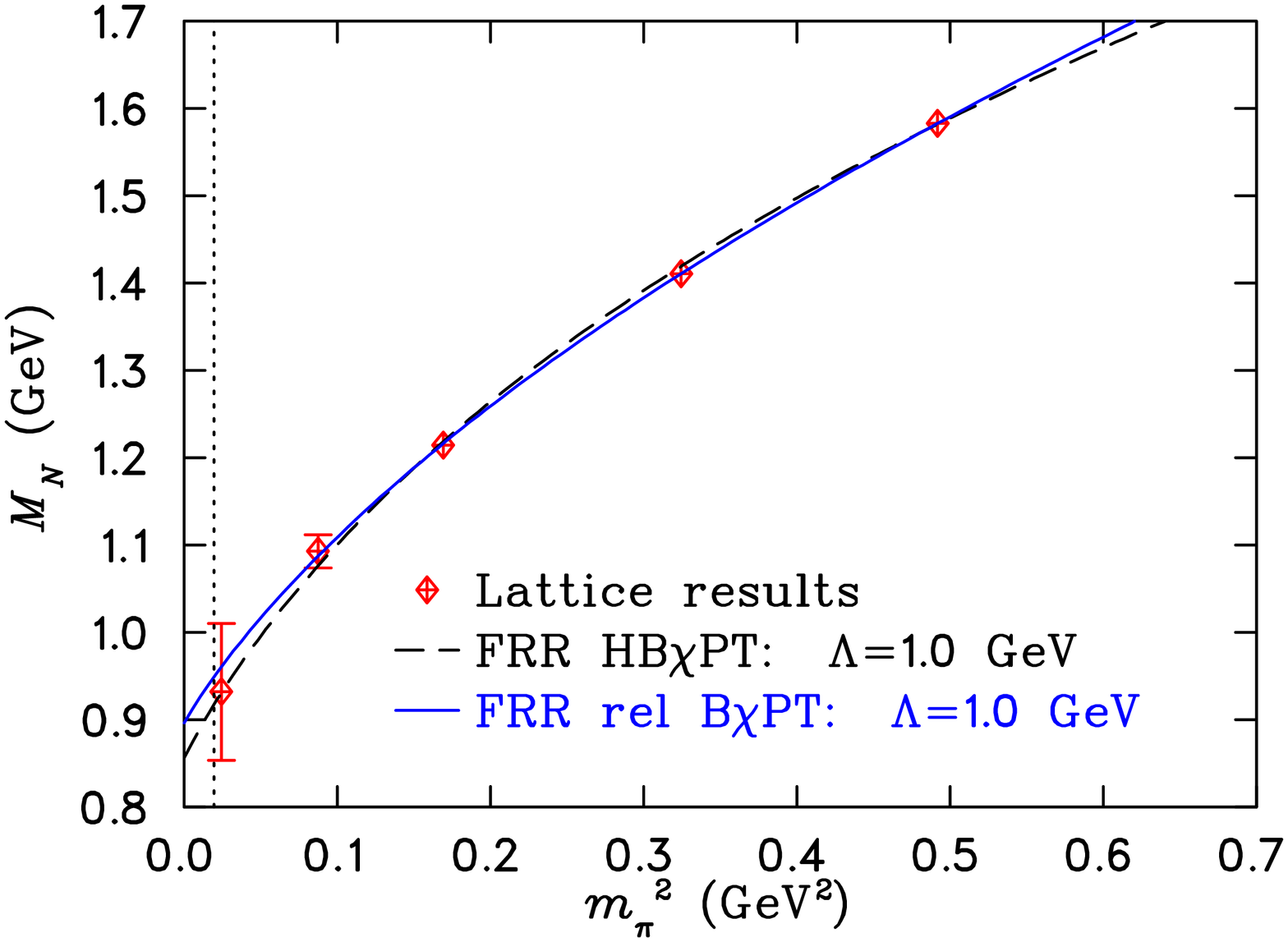}
\includegraphics[height=0.33\hsize,angle=90]{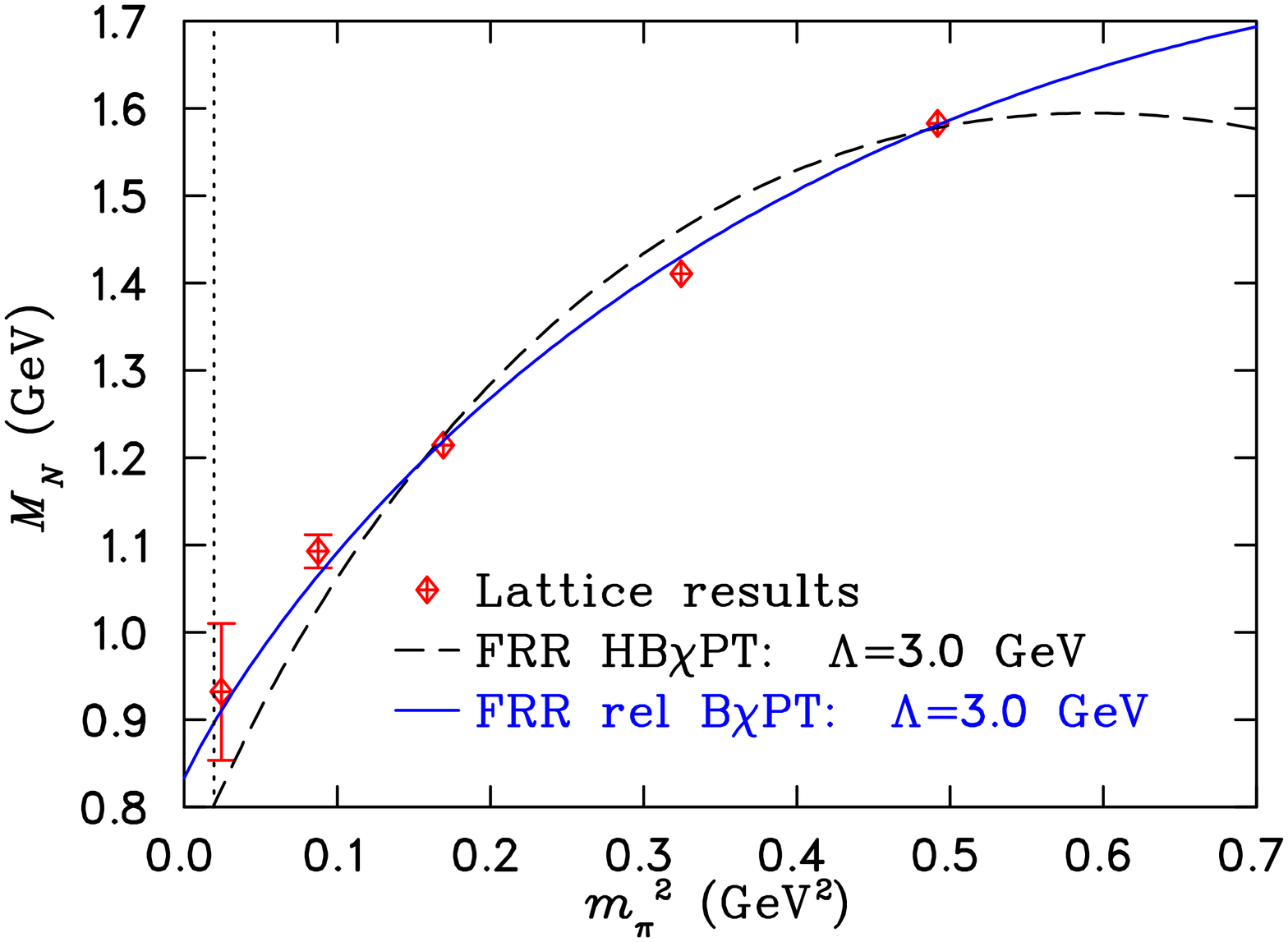}
\caption{(color online). Comparison of chiral extrapolations of the nucleon mass for HB$\chi$PT (black dashed curves) compared to the relativistic formula of B$\chi$PT (blue solid curves) from Eq.~(\ref{eqn:MNfitrel}), at various values of sharp-cutoff scale of $\La$. The extrapolation based on PACS-CS results \cite{Aoki:2008sm}, box size: $2.9$ fm. 
}
\label{fig:Aokiextlam}
\end{center}
\end{figure}

\begin{figure}[t]
\begin{center}
\includegraphics[height=0.33\hsize,angle=90]{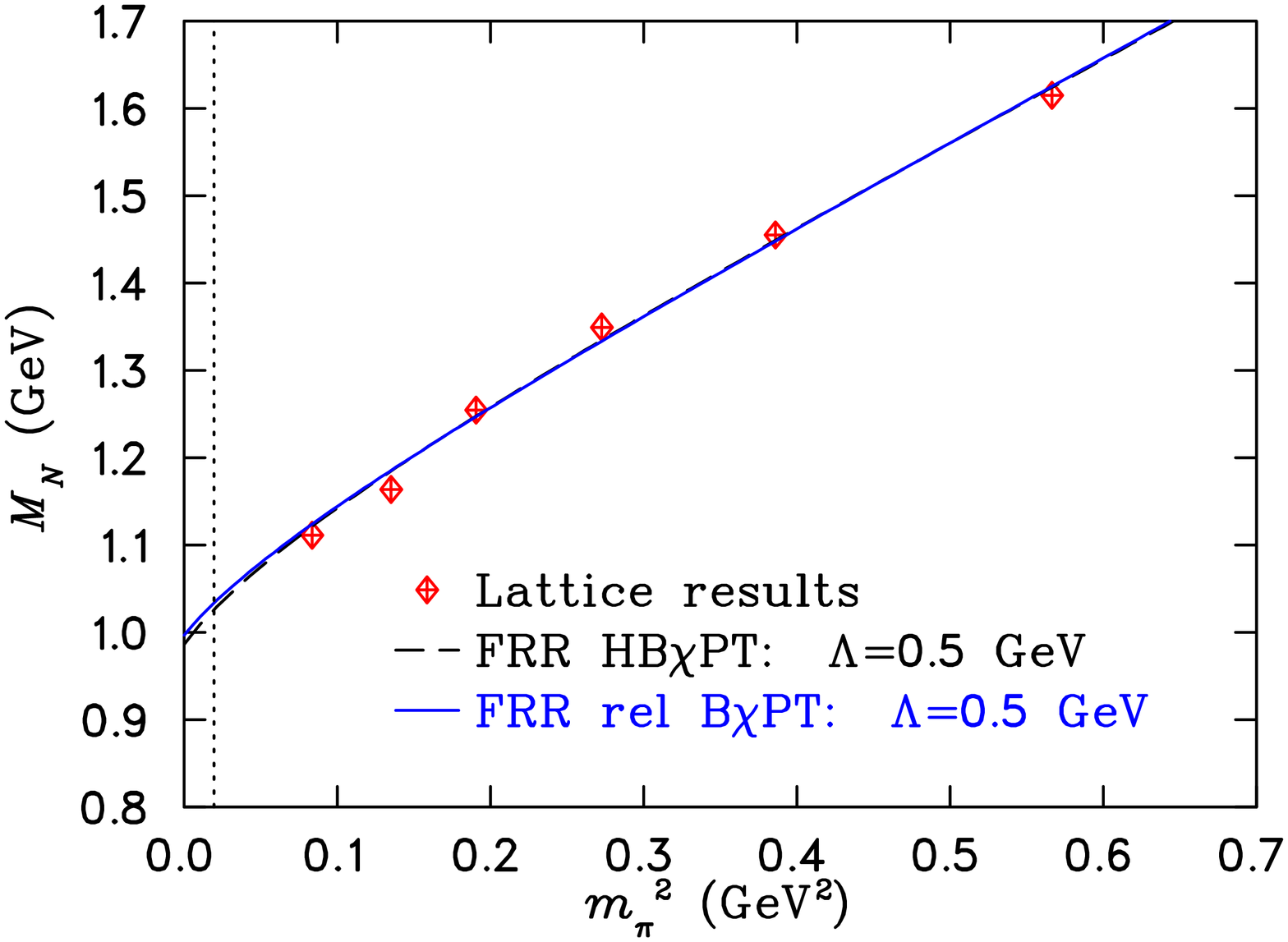}
\includegraphics[height=0.33\hsize,angle=90]{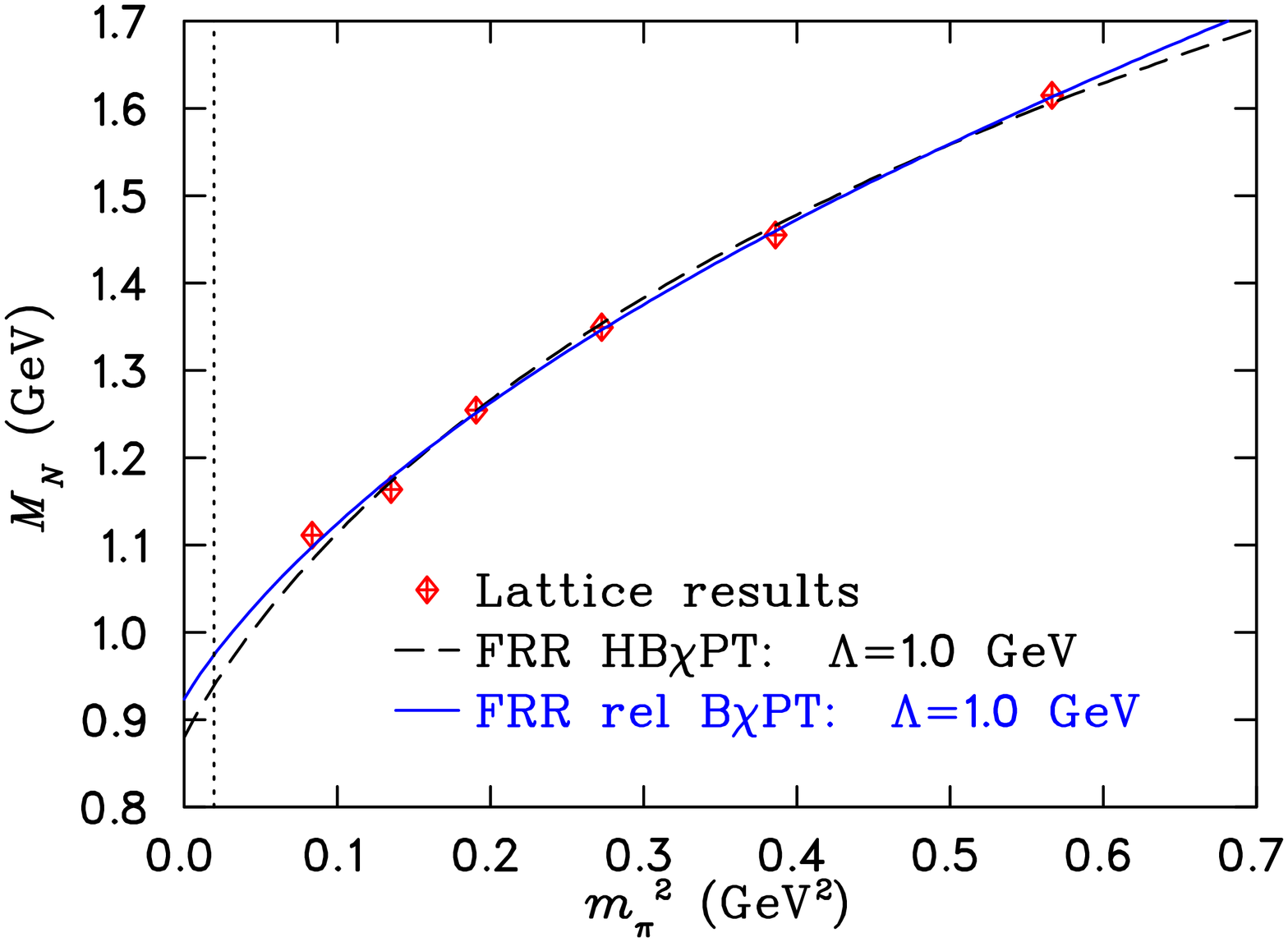}
\includegraphics[height=0.33\hsize,angle=90]{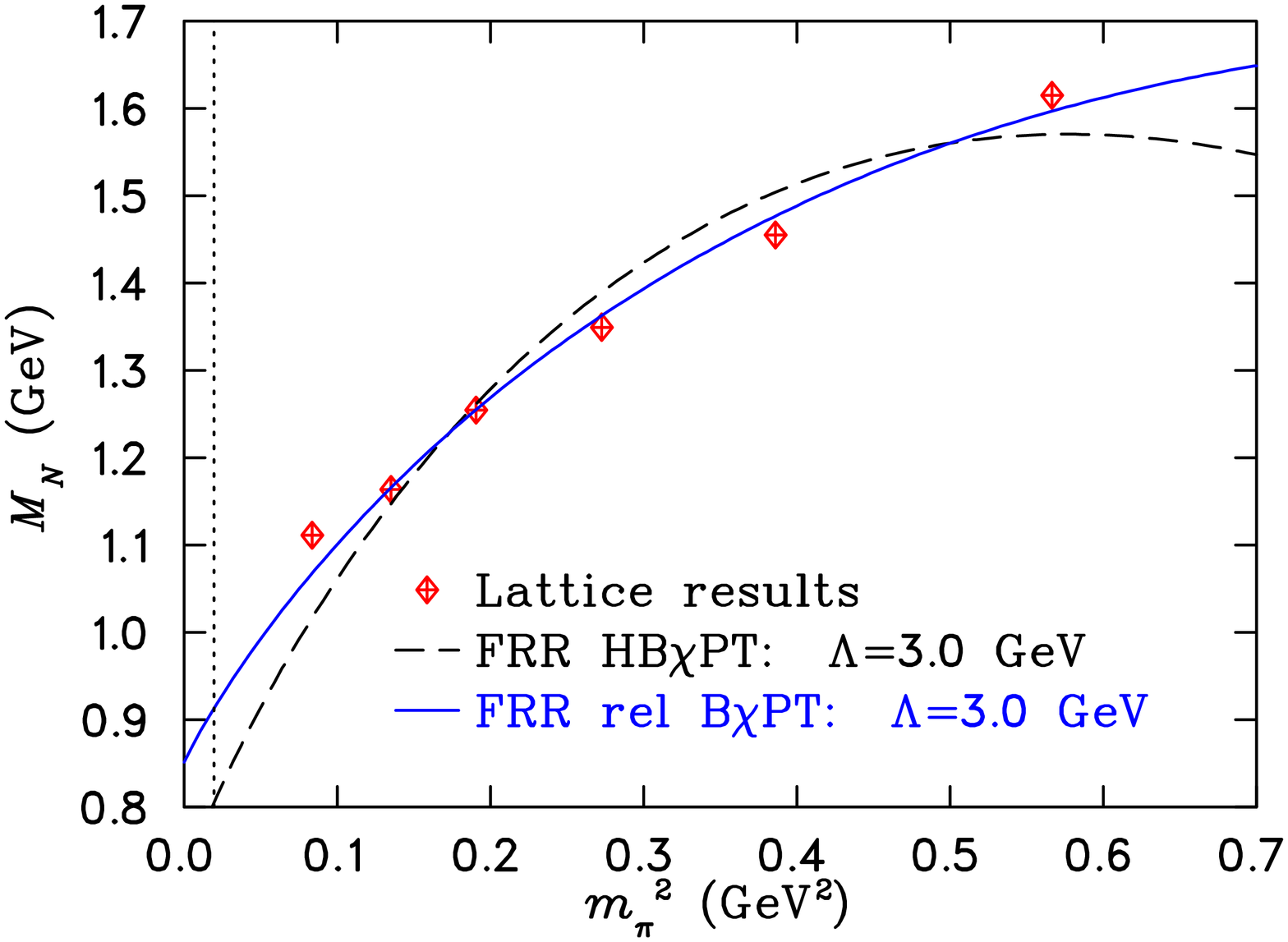}
\caption{(color online). Comparison of chiral extrapolations of the nucleon mass for HB$\chi$PT (black dashed curves) compared to the relativistic formula of B$\chi$PT (blue solid curves) from Eq.~(\ref{eqn:MNfitrel}), at various values of sharp-cutoff scale of $\La$. The extrapolation based on JLQCD results \cite{Ohki:2008ff}, box size: $1.9$ fm. 
}
\label{fig:extlam}
\end{center}
\end{figure}


\begin{figure}[h]
\begin{center}
\includegraphics[height=0.33\hsize,angle=90]{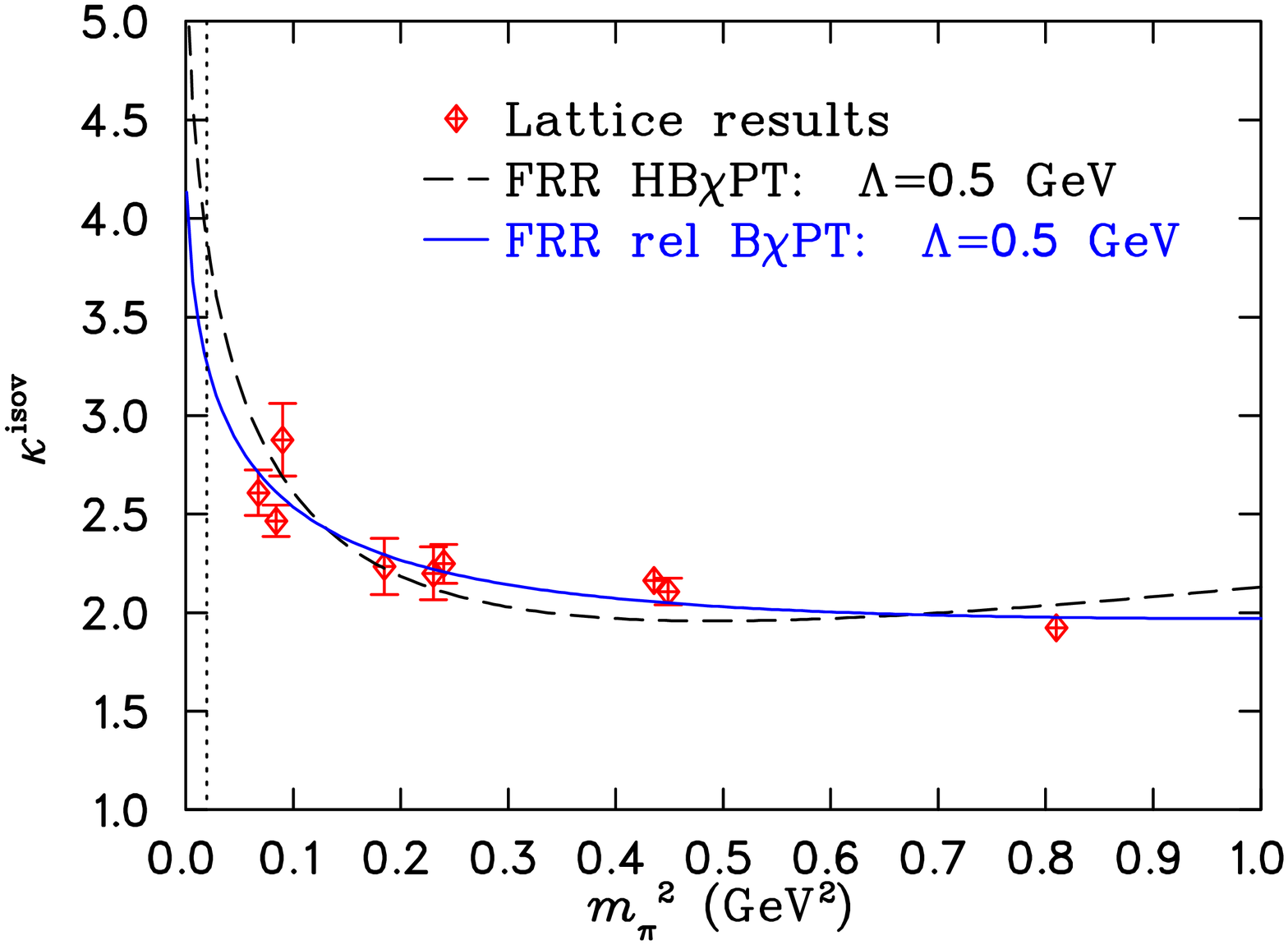}
\includegraphics[height=0.33\hsize,angle=90]{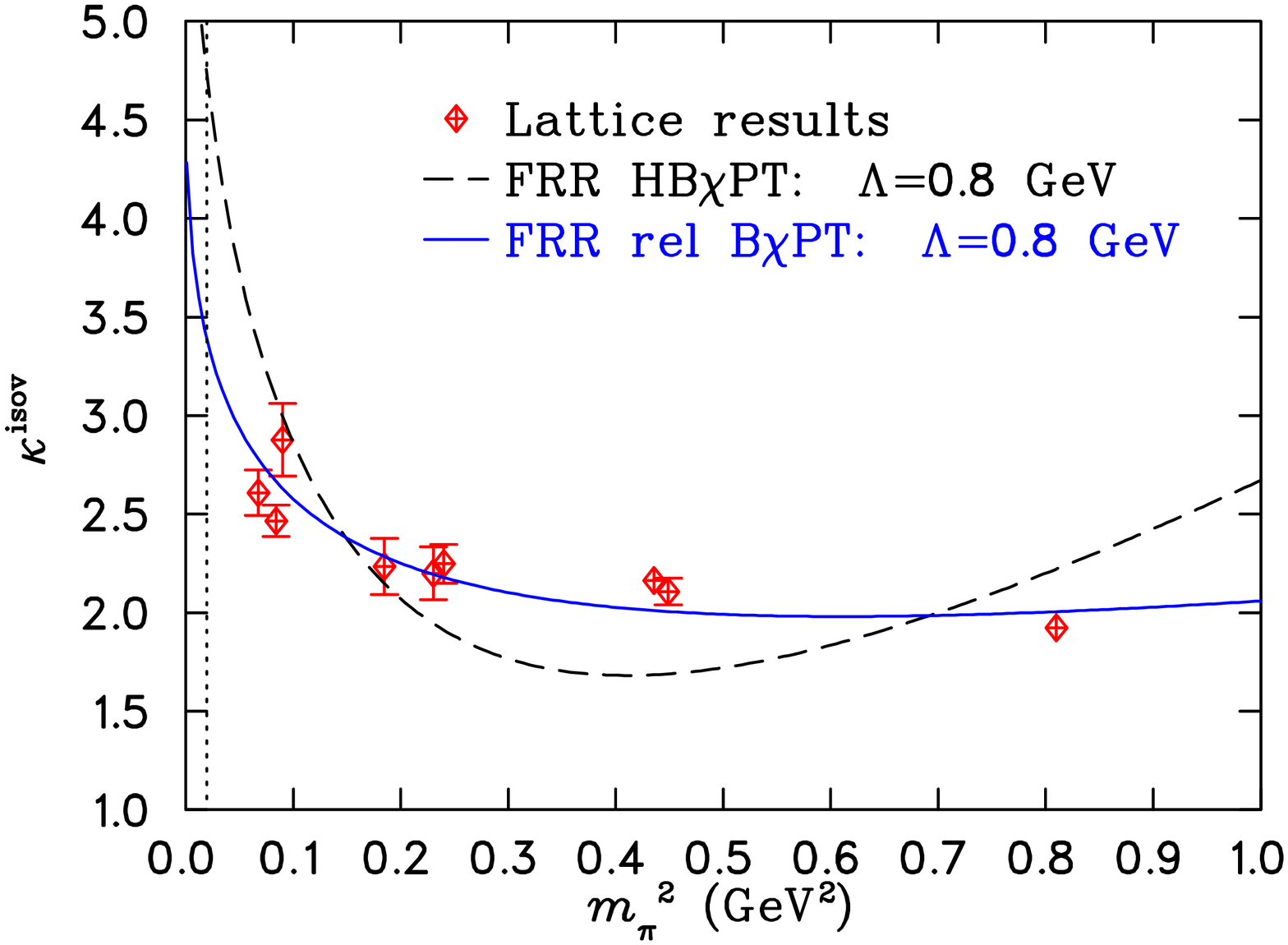}
\includegraphics[height=0.33\hsize,angle=90]{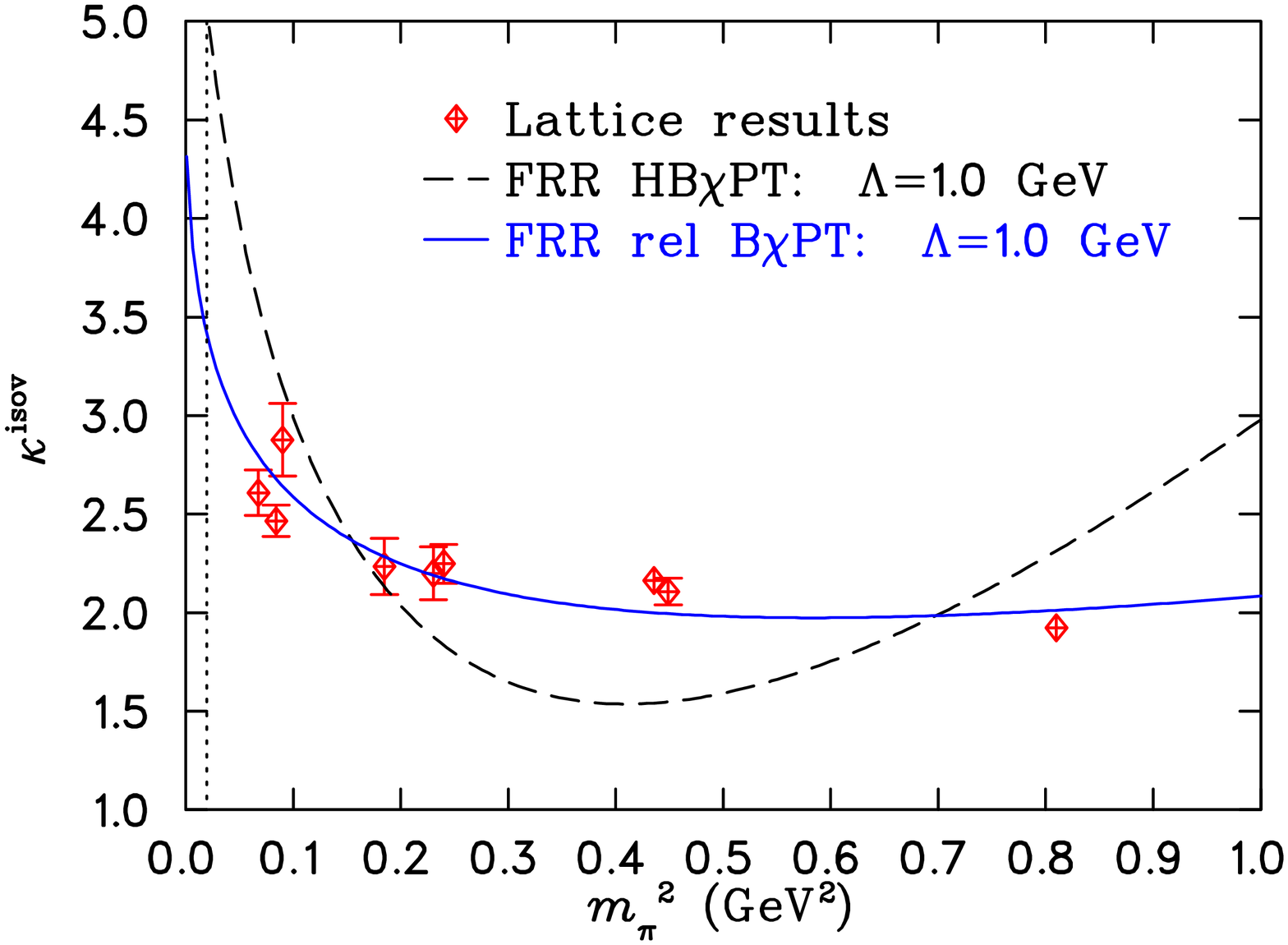}
\caption{(color online). Comparison of chiral extrapolations of the anomalous magnetic moment for HB$\chi$PT (black dashed curves) compared to the relativistic formula of B$\chi$PT (blue solid curves), for three different values of $\La$. The extrapolation based on QCDSF results \cite{Collins:2011mk}, box size: $1.7-2.9$ fm. 
}
\label{fig:kextlam}
\end{center}
\end{figure}
\end{widetext}

Previous studies in HB$\chi$PT have indicated that,
 for lattice results extending outside the 
chiral power-counting regime, the optimal value of the 
FRR scale $\La$  
is of the order $\La \approx 1$ GeV \cite{Hall:2010ai}. Clearly,  
Fig.\ \ref{fig:Aokiextlam} shows agreement 
that the best heavy-baryon result is obtained for $\La \approx 1$ GeV.  
The heavy-baryon extrapolation 
is much more sensitive to changes in the FRR scale $\La$ compared with  
the B$\chi$PT extrapolation, in agreement with Fig.~\ref{fig:LaDep}.

For small values of $\La$, the HB$\chi$PT and B$\chi$PT  
results are similar, since the chiral loops are suppressed.  
An almost-linear fit eventuates in both cases. 
This is not ideal, as Fig.~\ref{fig:Aokiextlam} indicates that 
neglecting the chiral curvature leads to a poor fit of the low-energy 
lattice results.  
For larger values of $\La$, the heavy-baryon extrapolation struggles 
to fit the lattice results due to large curvature in the heavy 
pion-mass region.  
The relativistic extrapolation appears to produce a more stable fit to the 
lattice results across a range of values of cutoff scale $\La$.

The importance of accommodating  
chiral curvature is 
even greater in the case of observables with  lower-order 
leading nonanalytic terms in their chiral expansions, such as the 
magnetic moment of the nucleon.
In the case of the AMMs, recent lattice QCD simulations by 
the QCDSF collaboration are used \cite{Collins:2011mk}.

 The results from QCDSF
were generated using $N_f = 2$ and the $\ca{O}(a)$-improved
 Wilson quark action, with box sizes ranging from $1.7$ to $2.9$ fm. 
To ensure that the lattice results from QCDSF
give a reasonable approximation to the infinite volume limit, 
the following restrictions are applied: $L>1.5$ fm and $m_\pi L > 3$. 
There are nine lattice points that satisfy these 
criteria from the original set of results.  
Additionally, the isovector combination of the nucleon 
($p - n$) is used in lattice QCD to avoid calculating the 
disconnected loops that occur in full QCD, which are computationally 
intensive, since they 
involve the calculation of all-to-all propagators.
In general,  diagrams contributing to the AMM   
of a hadron include photons coupling to 
sea-quark loops.
 In the special case of the isovector, the diagrams that include these 
disconnected loops cancel.

\begin{table}[t]
 \caption{A comparison of the values of the fit parameters $\Mc_N$ and $\cc_1$ for HB$\chi$PT and relativistic B$\chi$PT for various values of regularization scale $\La$, based on lattice results from PACS-CS.}
  \newcommand\T{\rule{0}{2.8ex}}
  \newcommand\B{\rule[-1.4ex]{0pt}{0pt}}
  \begin{center}
    \begin{tabular}{lllll}
      \hline
      \hline
      $\La$(GeV) &  $\McHB$(GeV) &  $\McB$(GeV) & $\ccHB$(GeV$^{-1}$) & $\ccB$(GeV$^{-1}$) \\
     \hline
$0.5$ & $0.953$($22$) & $0.963$($22$) & $-0.709$($9$) & $-0.660$($9$) \\
$1.0$ & $0.856$($22$) & $0.896$($22$) & $-0.970$($9$) & $-0.840$($9$) \\
$3.0$ & $0.717$($22$) & $0.833$($22$) & $-1.278$($9$) & $-0.981$($9$) \\
      \hline
    \end{tabular}
  \end{center}
\vspace{-6pt}
  \label{table:MNfitparamsPACS-CS}
\end{table}
\begin{table}[h]
 \caption{A comparison of the values of the fit parameters $\Mc_N$ and $\cc_1$ for HB$\chi$PT and relativistic B$\chi$PT for various values of regularization scale $\La$, based on lattice results from JLQCD.}
  \newcommand\T{\rule{0pt}{2.8ex}}
  \newcommand\B{\rule[-1.4ex]{0pt}{0pt}}
  \begin{center}
    \begin{tabular}{lllll}
      \hline
      \hline
       \T\B 
      $\La$(GeV) &  $\McHB$(GeV) &  $\McB$(GeV) & $\ccHB$(GeV$^{-1}$) & $\ccB$(GeV$^{-1}$) \\
     \hline
$0.5$ & $0.986$($8$) & $0.997$($8$) & $-0.676$($1$) & $-0.627$($1$) \\
$1.0$ & $0.880$($8$) & $0.924$($8$) & $-0.943$($1$) & $-0.811$($1$) \\
$3.0$ & $0.721$($8$) & $0.851$($8$) & $-1.265$($1$) & $-0.959$($1$) \\
      \hline
    \end{tabular}
  \end{center}
\vspace{-6pt}
  \label{table:MNfitparamsJLQCD}
\end{table}

The B$\chi$PT integral results from Eqs.~(\ref{eqn:frrtildeprot}) 
\& ~(\ref{eqn:frrtildeneut}) can be adapted for chiral extrapolation 
of the nucleon isovector 
by taking the difference between the proton 
and neutron AMM formulae. In addition, a linear term in $m_\pi^2$ is added 
with a free fit parameter $a_2$, which plays the role of compensating
 some of the high-momentum contributions:  
\eqb
\kappa_\ro{isov}(m_\pi^2,\La^2) = \kappa_p(m_\pi^2,\La^2) - 
\kappa_n(m_\pi^2,\La^2)+ a_2 m_\pi^2.
\eqe
Without it, the heavy-baryon result is just a straight line, and the
differences between the two frameworks are irreconcilable in this range
of pion masses.

The corresponding chiral behavior of the isovector nucleon 
 AMM
 is shown in Fig.\ \ref{fig:kextlam}  for three different values of $\La$, 
with corresponding values of the fit parameters 
shown in Table~\ref{table:AMMfitparams}. 
 Note that the curvature of the extrapolation 
using a sharp-cutoff regulator with $\La = 1.0$ GeV is already large. 
This is a consequence of the leading-order nonanalytic behavior of the 
AMM occurring at a lower chiral order ($\sim \!m_\pi$) than for  
 the nucleon mass ($\sim \!m_\pi^3$).  However, the extrapolation 
using the relativistic B$\chi$PT formulae of Eqs.\ (\ref{eqn:frrtildeprot})
\& (\ref{eqn:frrtildeneut}) is comparatively insensitive to changes in 
the FRR scale $\La$. This indicates that the B$\chi$PT formulae are largely
independent of the ultraviolet behavior. 

\begin{table}[t]
 \caption{A comparison of the values of the fit parameter $\kc_{\ro{isov}}$ and $a_2$ for HB$\chi$PT and relativistic B$\chi$PT for various values of regularization scale $\La$.}
  \newcommand\T{\rule{0pt}{2.8ex}}
  \newcommand\B{\rule[-1.4ex]{0pt}{0pt}}
  \begin{center}
    \begin{tabular}{lllll}
      \hline
      \hline
       \T\B 
      $\La$(GeV) &  $\kcHB$ & $\kcB$ & $a_2^{\ro{HB}}$ (GeV$^{-2}$) & $a_2^{\ro{B}}$ (GeV$^{-2}$) \\
     \hline
$0.5$ & $5.23$($5$)   &$4.13$($5$) & $0.70$($10$) & $0.14$($10$)\\
$0.8$ & $6.27$($5$)  & $4.28$($5$) & $3.42$($10$) & $0.65$($10$)\\
$1.0$ & $6.68$($5$)  & $4.31$($5$) & $5.46$($10$) & $0.83$($10$)\\
      \hline
    \end{tabular}
  \end{center}
\vspace{-6pt}
  \label{table:AMMfitparams}
\end{table}

As shown in the previous section the naturalness problem doesn't arise in the AMM, in the same drastic way as it does in the polarizability. However for a broader range of pion masses the problem starts to show.
In the case of AMM, $a_2^{HB}$ quickly becomes larger with increasing $\Lambda$, in order to accommodate the large higher-momentum contributions.

\section{Conclusion: when heavy-baryon fails}
\label{sect:conc}
The HB$\chi$PT  and B$\chi$PT can be viewed as two different ways 
of organizing the chiral EFT expansion in the baryon sector. While 
the heavy-baryon expansion
is often considered to be more consistent from the power-counting point of 
view, it appears to be less natural. 
Certain terms that are nominally 
suppressed by powers of $m_\pi/M_N$, and hence dropped in HB$\chi$PT as
being `higher order', appear to be significant in explicit calculations.   

The problem is more pronounced in some quantities and less in others.
To quantify this, one needs to note the power
of the expansion parameter at which the chiral loops begin to contribute
to the quantity in question.
For the considered examples of the nucleon mass, AMMs, and polarizability, 
this power index is $3$, $1$, and $-1$, respectively. 
The smaller the index, the
greater is the 
 difficulty for HB$\chi$PT to describe this quantity in a natural way. 
For quantities
with a negative index, a dramatic failure of HB$\chi$PT is expected. 

The negative index simply means that the chiral expansion of that quantity 
begins at lowest order 
with negative powers of $m_\pi$. Apart from polarizabilities, 
the most notable quantities of this kind are the coefficients of the 
effective-range expansion of the nuclear force. As is known, 
the non-relativistic
$\chi$PT in the two-nucleon sector \cite{Kaplan:1998we} 
failed to describe these 
quantities \cite{Cohen:1998jr}, 
thus precluding the idea of `perturbative pions' in this sector   
\footnote{
Recent work by Epelbaum and Gegelia \cite{Epelbaum:2012ua}, 
appearing after the submission of this paper, suggests an alternative method in the two-nucleon sector, also based on a covariant approach without the heavy-baryon expansion.}.
The present work encourages one to think that B$\chi$PT can 
solve this problem, as is 
 the case for nucleon 
polarizabilities.

It certainly is important to understand the origin of the
 apparent difficulty of the HB expansion in estimating
 quantum corrections in certain
observables, at least in a way it has been understood for the scalar 
form factor~\cite{Becher:1999he}. Here, it was only shown that the behaviour  
comes from the ``soft'' momentum region, 
and  a criterion for determining the region relevant to this problem was 
conjectured. 
The understanding of the origin presents a challenge for future studies.

\begin{acknowledgments} 
V.P. is thankful to Dr.\ Nikolai Kivel for stimulating discussions of
the heavy-quark and heavy-baryon expansions. This work has been supported by the Deutsche Forschungsgemeinschaft through the Collaborative Research Centre SFB1044, and by the Australian Research Council through grant DP110101265. 
\end{acknowledgments}

\appendix
\section{The quark-mass dispersion relation and subtractions}
\label{sect:subt}
According to the Gell-Mann--Oakes--Renner relation (GOR)
\cite{GellMann:1968rz},
$m_\pi^2 \propto m_q$ for a light-quark mass $m_q$. Thus, 
the pion-mass dispersion relation can be translated into 
a quark-mass dispersion relation (as seen in Eq.~(\ref{eqn:disprel})):
\eqb
 f(m_q) = -\f{1}{\pi}\!\int\limits_{-\infty}^0\!\ro{d}t \, 
\f{\ro{Im} f(t)}{t-m_q}\, ,
\eqe
with $f$ being a function of $m_q$.  
The above allusion to GOR implies that this relation
is valid for small $m_q$ only. However, if its validity were assumed for
all $m_q$, the issue of its convergence for a given
quantity $f$ could be considered as follows.  

The unsubtracted
dispersion relation implies that $f~\propto~1/m_q$, for large $m_q$,
which obviously cannot be true for every quantity. For example, 
In the case of the nucleon mass $f\equiv M_N$, 
it is expected  
that  $f \propto m_q$ for large $m_q$.
Therefore, the relation needs to be subtracted at least twice, i.e:
\eqb
\label{eqn:disprel4a}
 M_N(m_q)  =  \, \Mc_N  + \, a_1 m_q   
-\f{m_q^2}{\pi}\!\int\limits_{-\infty}^0\! \frac{\ro{d}t}{t^2}\,
\f{\ro{Im} \,M_N(t)}{t-m_q}\, ,
\eqe
with $\Mc_N$ and $a_1$ being the subtraction constants. 

In another example, the nucleon AMM should behave a constant for large 
$m_q$ and hence one subtraction will suffice:
\eqb
\label{eqn:disprel4b}
 \kappa_N(m_q)  = \,\,\stackrel{\circ}{\kappa}_N   
-\,\f{m_q}{\pi}\!\int\limits_{-\infty}^0\! \frac{\ro{d}t}{t}\,
\f{\ro{Im} \, \kappa_N(t)}{t-m_q}\, ,
\eqe
where $\stackrel{\circ}{\kappa}_N$ is the AMM in the chiral limit,
playing the role of the subtractions constant.
 
The sufficiency of these subtractions is confirmed in leading-order 
$\chi$PT calculations. 
At higher chiral orders,  
more subtractions are needed as new low-energy
 constants arise to play the role of the subtraction constants. 
The above analysis determines only the minimal number of
subtractions for a given quantity.


\bibliographystyle{apsrev}
\bibliography{ref_disp}

\end{document}